\documentclass[12pt,preprint]{aastex}

\shorttitle{eruptive character of large solar flares}
\shortauthors{Li et al.}

\begin{document}

\title{Magnetic Flux of Active Regions Determining the Eruptive Character of Large Solar Flares}

\author{Ting Li\altaffilmark{1,2}, Yijun Hou\altaffilmark{1,2}, Shuhong Yang\altaffilmark{1,2}, Jun Zhang\altaffilmark{3,1}, Lijuan Liu\altaffilmark{4,5} \& Astrid
M. Veronig\altaffilmark{6}}

\altaffiltext{1}{CAS Key Laboratory of Solar Activity, National
Astronomical Observatories, Chinese Academy of Sciences, Beijing
100101, China; liting@nao.cas.cn} \altaffiltext{2}{School of
Astronomy and Space Science, University of Chinese Academy of
Sciences, Beijing 100049, China} \altaffiltext{3}{School of Physics
and Materials Science, Anhui University, Hefei 230601, China}
\altaffiltext{4}{School of Atmospheric Sciences, Sun Yat-sen
University, Zhuhai, Guangdong, 519082, China} \altaffiltext{5}{CAS
Center for Excellence in Comparative Planetology, China}
\altaffiltext{6}{Institute of Physics \& Kanzelh\"ohe Observatory
for Solar and Environmental Research, University of Graz, A-8010
Graz, Austria}

\begin{abstract}

We establish the largest eruptive/confined flare database to date
and analyze 322 flares of \emph{GOES} class M1.0 and larger that
occurred during 2010$-$2019, i.e., almost spanning the entire solar
cycle 24. We find that the total unsigned magnetic flux
($\Phi$$_{AR}$) of active regions (ARs) is a key parameter in
governing the eruptive character of large flares, with the
proportion of eruptive flares exhibiting a strong anti-correlation
with $\Phi$$_{AR}$. This means that an AR containing a large
magnetic flux has a lower probability for the large flares it
produces to be associated with a coronal mass ejection (CME). This
finding is supported by the high positive correlation we obtained
between the critical decay index height and $\Phi$$_{AR}$, implying
that ARs with a larger $\Phi$$_{AR}$ have a stronger magnetic
confinement. Moreover, the confined flares originating from ARs
larger than 1.0$\times$$10^{23}$ Mx have several characteristics in
common: stable filament, slipping magnetic reconnection and strongly
sheared post-flare loops. Our findings reveal new relations between
the magnetic flux of ARs and the occurrence of CMEs in association
with large flares. These relations obtained here provide
quantitative criteria for forecasting CMEs and adverse space
weather, and have also important implications for ``superflares" on
solar-type stars and stellar CMEs. The link of database
https://doi.org/10.12149/101030

\end{abstract}


\section{Introduction}

Solar flares and coronal mass ejections (CMEs) are the most
energetic phenomena in our solar system and are the dominant
contributors to adverse space weather at Earth (Gosling et al. 1991;
Green et al. 2018). They originate from the rapid release of free
magnetic energy stored in the sheared or twisted magnetic fields of
active regions (ARs) through magnetic reconnection (Forbes 2000;
Shibata \& Magara 2011). Magnetic reconnection is believed to be a
fundamental process in magnetized plasma systems throughout the
Universe where magnetic energy is stored over relatively long times
to be released suddenly in bursts of various forms (thermal,
kinetic, and high-energy particle) (Priest \& Forbes 2000; Su et al.
2013). Flares associated with a CME are usually referred to as
eruptive events, while flares that are not accompanied by a CME are
called confined or ``CME-less" events (Svestka 1986; Moore et al.
2001). The association rate of flares and CMEs has revealed that
most small flares occur without a CME, whereas for large flares
(M-class, X-class) the CME-association is steeply increasing, and
reaches 100\% for the biggest events (Andrews 2003; Yashiro et al.
2006). The broad variety of strong space weather effects is mostly
due to the CME rather than the flare itself. Our understanding of
the physical mechanism of flares and their relationship with CMEs is
important to forecast space weather in the near-Earth environment
(Forbes 2000; Shibata \& Magara 2011). Meanwhile, the solar
flare-CME paradigm might be applied to magnetic activities in other
stars, which is vital for the question of exoplanet habitability and
the evolution of stellar mass loss and rotation (Khodachenko et al.
2007; Lammer et al. 2007).

Substantial observational studies have revealed that a flare would
be confined if the strapping magnetic field overlying the flaring
region is too strong or does not decrease sufficiently fast with
height (Green et al. 2002; Wang \& Zhang 2007; Cheng et al. 2011;
Yang et al. 2014; Chen et al. 2015; Thalmann et al. 2015). To
quantify the decline of the strapping field with height, the decay
index has been used (Kliem \& T{\"o}r{\"o}k 2006; Fan \& Gibson
2007; Zuccarello et al. 2015), i.e., $\emph{n}$=--$\emph{d}$
$\ln$$B_{hor}$/$\emph{d}$ $\ln$$\emph{h}$, with $B_{hor}$ denoting
the horizontal field and $\emph{h}$ the height in the corona. The
torus instability of a magnetic flux rope occurs when the decay
index $\emph{n}$ reaches a critical value $n_{crit}$$\approx$1.5
(Kliem \& T{\"o}r{\"o}k 2006; Fan \& Gibson 2007). Recent
magnetohydrodynamics (MHD) simulations showed that the overlying
field lines form a confining cage and a weaker magnetic cage would
produce a more energetic eruption with a CME (Amari et al. 2018).
Another important factor that governs the eruptive character of
solar flares is the non-potentiality of ARs. Statistical studies
have shown that flare and CME productivity are correlated with
magnetic shear, electric current, magnetic free energy, etc (Hagyard
\& Rabin 1986; Falconer et al. 2002; Liu et al. 2016a). It is
suggested that AR eruptivity is related to the relative value of
magnetic non-potentiality over the restriction of the background
field (Sun et al. 2015).

In this paper, we derive important quantitative relations between
the magnetic properties of ARs and the eruptive character of large
solar flares, based on the \emph{Solar Dynamics Observatory}
(\emph{SDO}; Pesnell et al. 2012) observations during the period of
solar cycle 24. A total of 322 flares (170 eruptive and 152
confined) of Geostationary Operational Environmental Satellite
(\emph{GOES}) class M1.0 and larger that occurred within
45$^{\circ}$ from the central meridian, from June 2010 until June
2019, are selected. About 51\% (155 of 301) of the M-class flares
are eruptive and the percentage increases up to $\sim$71\% (15 of
21) for X-class flares, similar to the previous results (Yashiro et
al. 2006). To our knowledge, the eruptive/confined flare sample
established in this study is by far the largest one in the era of
\emph{SDO}. We show that total unsigned magnetic flux of ARs
($\Phi$$_{AR}$) is a decisive parameter in governing the eruptive
character of flares, and the proportion of eruptive flares exhibits
a strong anti-correlation with $\Phi$$_{AR}$. This finding is
further supported by the high correlation obtained between
$\Phi$$_{AR}$ and the critical height for torus instability.

The rest of the paper is organized as follows. In Sections 2 and 3,
we describe the data analysis and show the statistical results,
respectively. Section 4 presents the detailed analysis for six
events as typical examples. Finally, we summarize our findings and
discuss the implications in Section 5.

\section{Data Analysis}
\subsection{Event Sample}

The \emph{SDO} satellite has already provided a rich database since
its launch in February 2010. Until now, its observation period lasts
about 10 years and almost spans the entire solar cycle 24. Thus it
is a good opportunity to carry out statistical analysis about the
flare-CME mechanism based on the \emph{SDO} observations. Firstly,
we examined a database RibbonDB presented by Kazachenko et al.
(2017) and selected all 302 flare events larger than M1.0 that
occurred within $45^{\circ}$ from the central meridian, from June
2010 until April 2016. To extend the time period, we checked for the
\emph{GOES} soft X-ray (SXR) flare catalog to search for flare
events from May 2016 to June 2019 and found 20 flares of GOES class
M1.0 and greater. A total of 322 flare events are involved in our
database over a nine-year period (see Table FlareM1.0). Secondly,
for each flare, its CME association was determined by checking the
CME catalog\footnote{\url{https://cdaw.gsfc.nasa.gov/CME\_list/}}
(Gopalswamy et al. 2009) of the Solar and Heliospheric Observatory
(SOHO)/Large Angle and Spectrometric Coronagraph (LASCO). We
regarded a flare as eruptive if the CME onset time was within 90 min
of the flare start time and the position angle of the CME agreed
with the quadrant on the Sun where the flare occurred. Moreover, we
also inspected the observations of the \emph{twin Solar Terrestrial
Relations Observatory} (\emph{STEREO}; Kaiser et al. 2008; Howard et
al. 2008) to check from a different viewing angle if there is an
associated CME. For the events difficult to determine the
classification, e.g., there are two flares within a short time or
the CME is too weak to be identified, we then checked the EUV
observations from the Atmospheric Imaging Assembly (AIA; Lemen et
al. 2012) on board the \emph{SDO} and identified the coronal EUV
wave manually. If a global coronal EUV wave was associated with the
flare, the flare was classified as eruptive. Out of these 322
flares, 170 ($\sim$53\%) events were eruptive (155 M- and 15 X-) and
152 ($\sim$47\%) were confined (146 M- and 6 X-).

\subsection{Data and Methods}

We investigated the relations between the AR parameters (unsigned AR
magnetic flux $\Phi$$_{AR}$ and AR area) and the eruptive character
of large solar flares. The AR parameters in RibbonDB catalog
(Kazachenko et al. 2017) are calculated based on the full-disk
Helioseismic and Magnetic Imager (HMI; Scherrer et al. 2012) vector
magnetogram data series (\verb#hmi.B_720s#) before the flare onset
time. To avoid noisy magnetic fields, only pixels that host a normal
component of the magnetic field $|$$B_{n}$$|$$>$100 G are
considered. For the flare events that are not included in the
RibbonDB catalog, we use the available vector magnetograms
(\verb#hmi.sharp_cea_720s#) from Space-Weather HMI AR Patches (Bobra
et al. 2014) before the flare onset. The magnetograms were re-mapped
using a cylindrical equal area projection with a pixel size of
$\sim$0$\arcsec$.5 and presented as (B$_{r}$, B$_{\theta}$,
B$_{\phi}$) in heliocentric spherical coordinates corresponding to
(B$_{z}$, -B$_{y}$, B$_{x}$) in heliographic coordinates (Sun 2013).
Similarly, to calculate the AR magnetic flux and AR area, we
consider all pixels of $|$$B_{r}$$|$$>$100 G. Moreover, RibbonDB
catalog (Kazachenko et al. 2017) also includes the parameters of
flare ribbons such as the flare ribbon reconnection flux
$\Phi$$_{ribbon}$, the cumulative flare ribbon area S$_{ribbon}$,
the ratio of the AR magnetic flux involved in the flare reconnection
R$_{flux}$ ($\Phi$$_{ribbon}$/$\Phi$$_{AR}$) and the area ratio
R$_{S}$ (S$_{ribbon}$/S$_{AR}$). In this work, we also use these
parameters of 302 flare events larger than M1.0 and investigate
their distributions and correlations in eruptive and confined
flares.

The role of the background coronal fields in confined and eruptive
flares was estimated by calculating the decay index n above the ARs.
In order to extrapolate the 3D magnetic field in the entire AR
volume, we use the Fourier transformation method (Alissandrakis
1981) to extrapolate the potential field. The method yields the
local potential field with a resolution around 0.72 Mm, same as the
resolution of the boundary condition. The boundary condition is the
normal component of the photospheric magnetic field from
Space-Weather HMI AR Patches (Bobra et al. 2014) observed prior to
the flare start. From the extrapolated field, the mean value of the
horizontal magnetic field, $<$$B_{hor}$$>$, as a function of height
is obtained along the main polarity inversion line (PIL) and an
average decay index, $<$n$>$, is then derived. Here, the main PIL
was identified as zero Gauss contour in the bottom vertical magnetic
field (B$_{r}$) image from the extrapolated potential fields
(Bokenkamp 2007; Vasantharaju et al. 2018). To analyze the structure
and dynamics of typical flare examples, we used the E(UV)
observations from the AIA, with a resolution of $\sim$0$\arcsec$.6
per pixel and a cadence of 12(24) s. Five channels of AIA 1600, 304,
171, 94 and 131 {\AA} were mainly applied to analyze the appearances
of the flares. The full-disk line-of-sight (LOS) magnetic field data
from the HMI are also used to present the ARs producing the typical
flare examples.

\section{Statistical Results}

\subsection{Magnetic Properties of ARs and Eruptive Character of Solar Flares}

Figure 1(a) shows the scatter plot of the flare peak X-ray flux
versus $\Phi$$_{AR}$. Blue (red) circles are the eruptive (confined)
flares. Obviously, when $\Phi$$_{AR}$ is small enough, the flares
tend to be eruptive (Area A in Figure 1(a)). About 92\% (36 of 39)
of events occurring in ARs with $\Phi$$_{AR}$ smaller than
3.0$\times$$10^{22}$ Mx are eruptive. An overwhelming majority of
flares that are hosted by ARs with a large magnetic flux do not
generate CMEs (Area C in Figure 1(a)). The proportion of confined
flares of GOES class $\geq$M1 is $\sim$93\% (26 of 28) corresponding
to the AR with $\Phi$$_{AR}$ larger than 1.0$\times$$10^{23}$ Mx. We
examined two special eruptive events (M4.0-class flare on 24 October
2014 in Figure 10 and X1.8-class event on 20 December 2014) in Area
C of Figure 1(a) and found that they either were located at the edge
of the AR or caused a sympathetic eruption of a nearby quiescent
filament. If the AR has a moderate magnetic flux (larger than
3.0$\times$$10^{22}$ Mx and smaller than 1.0$\times$$10^{23}$ Mx),
the likelihood of eruptive and confined events appears to be almost
equal (132 eruptive flares and 126 confined events of 258 in Area B
of Figure 1(a)). The scatter plot of the flare peak X-ray flux
versus total AR area shows a similar trend (Figure 1(b)). All flares
in ARs with an area smaller than 5.0$\times$$10^{19}$ $cm^{2}$ are
eruptive (Area A in Figure 1(b)) and all flares in ARs larger than
3.0$\times$$10^{20}$ $cm^{2}$ are confined (Area C in Figure 1(b)).

Figures 1(c)-(d) display the histograms for confined and eruptive
events. There are significant differences in distributions of AR
magnetic flux and AR area between the confined and eruptive cases.
The confined events have larger AR magnetic flux and AR area. The
averages of the log values of $\Phi$$_{AR}$ (indicated by vertical
dotted lines) are 6.3$\times$$10^{22}$ Mx and 4.4$\times$$10^{22}$
Mx for confined and eruptive cases, respectively. The log-mean
values of AR area for the confined and eruptive events are
1.5$\times$$10^{20}$ $cm^{2}$ and 1.2$\times$$10^{20}$ $cm^{2}$,
respectively. Based on the the number distributions of AR magnetic
flux and AR area between the confined and eruptive cases, we display
the relations of the proportions of eruptive flares P$_{E}$
(P$_{E}$=N$_{E}$/(N$_{E}$+N$_{C}$), N$_{E}$ and N$_{C}$ are numbers
of eruptive and confined events, respectively) with $\Phi$$_{AR}$
and AR area in Figures 1(e)-(f). It can be seen that P$_{E}$
decreases with $\Phi$$_{AR}$. The proportion P$_{E}$ has a strong
anti-correlation with $\Phi$$_{AR}$ at the Spearman rank order
correlation coefficient r$_{s}$ of $-$0.97. The Spearman rank
correlation provides a measure of the monotonic relationship between
two variables. The linear fitting to the scatter plot provides the
relation of

P$_{E}$=(-0.75$\pm$0.06)log$|$$\Phi$$_{AR}$$|$+(17.53$\pm$1.27), (1)

where $\Phi$$_{AR}$ is in units of [Mx].

Similarly, the proportion P$_{E}$ shows a strong anti-correlation
with AR area (r$_{s}$=$-$0.95), and provides the relation of

P$_{E}$=(-0.76$\pm$0.09)logS$_{AR}$+(15.70$\pm$1.86), (2)

where S$_{AR}$ is in units of [cm$^{2}$].

\subsection{Role of the Background Coronal Fields}

We investigate the role of the background coronal fields by
calculating the decay index n above the ARs. Figure 2 shows four
examples including one eruptive flare in Area A of Figure 1(a) and
three confined flares in Area C of Figure 1(a). Black asterisks
denote the $<$$B_{hor}$$>$ versus h profiles and blue diamonds are
the $<$n$>$ versus h profiles. The error bars mark the corresponding
standard deviation. The critical height for torus instability
$h_{crit}$ corresponds to the height where $<$n$>$ reaches a value
of 1.5 (Kliem \& T{\"o}r{\"o}k 2006; Fan \& Gibson 2007). Clearly,
the $h_{crit}$ value of AR 11305 ($\sim$ 17 Mm) with a small
magnetic flux is lower than those of three other ARs (36$-$60 Mm)
with larger magnetic fluxes, which means that the constraining field
above AR 11305 producing an eruptive flare decays more rapidly than
other ARs with confined flares, and therefore a perturbation in the
lower corona may cause the CME seed to erupt out more easily (Wang
\& Zhang 2007; Liu et al. 2018).

Following the procedure described above, we estimated the critical
decay index heights $h_{crit}$ for 82 events (including all the
events in Areas A and C of Figure 1(a) and 15 flares in Area B of
Figure 1(a)). Figure 3(a) shows the scatter plot of $h_{crit}$
versus $\Phi$$_{AR}$. It can be seen that $h_{crit}$ increases with
$\Phi$$_{AR}$. This indicates that ARs with a larger magnetic flux
tend to have stronger constraining field. The critical decay index
height has a strong correlation with AR magnetic flux at the
Spearman rank order correlation coefficient r$_{s}$ of 0.86. The
linear fitting to the scatter plot provides the relation of

$h_{crit}$=(38.31$\pm$2.37)log$|$$\Phi$$_{AR}$$|$+($-$834.53$\pm$53.92),
(3)

where $h_{crit}$ and $\Phi$$_{AR}$ are in units of [Mm] and [Mx],
respectively.

Using this equation, an $\Phi$$_{AR}$ value of 3.0$\times$$10^{22}$
Mx yields a $h_{crit}$ of $\sim$27 Mm (left vertical and bottom
horizontal lines in Figure 3(a)) and 1.0$\times$$10^{23}$ Mx
corresponds to $h_{crit}$ of about 47 Mm (right vertical and top
horizontal lines in Figure 3(a)). In Figure 3(b), we plot the flare
peak X-ray flux versus $h_{crit}$. All flares $\geq$M1 (28 events)
with a $h_{crit}$ value smaller than 27 Mm are eruptive (Area A in
Figure 3(b)), and about 95\% (20 of 21) of events with $h_{crit}$
larger than 47 Mm are confined (Area C in Figure 3(b)). The results
of Figures 1 and 3 suggest that stronger strapping fields over the
ARs with a larger magnetic flux play the major role in confining the
eruption.

\subsection{Relations of Flare Reconnection Flux with Flare Peak X-Ray Flux}

Figure 4 shows the scatter plots of flare ribbon reconnection flux
and cumulative flare ribbon area versus flare peak X-ray flux. We
find that flare reconnection flux $\Phi$$_{ribbon}$ correlates with
flare peak X-ray flux F$_{SXR}$ at a moderate rank order correlation
coefficient r$_{s}$ of 0.51 for all the flares (Figure 4(a)). By
fitting the data, we obtained their empirical relationship

log$|$$\Phi$$_{ribbon}$$|$=(0.51$\pm$0.04)logF$_{SXR}$+(24.02$\pm$0.17),
(4)

where $\Phi$$_{ribbon}$ and F$_{SXR}$ are in units of [Mx] and
[W/m$^{2}$], respectively.

The rank order correlation coefficient r$_{s}$ for the subset of
eruptive flares (r$_{s}$=0.58) is larger than r$_{s}$ for the
confined flares (r$_{s}$=0.42). The corresponding fitting parameters
for the subsets of confined and eruptive flares show no significant
differences.

The ribbon area and flare peak X-ray flux (Figure 4(b)) also show a
moderate correlation with a rank order correlation coefficient
r$_{s}$ of 0.58 and their relation is

logS$_{ribbon}$=(0.49$\pm$0.03)logF$_{SXR}$+(21.12$\pm$0.14), (5)

where S$_{ribbon}$ and F$_{SXR}$ are in units of [cm$^{2}$] and
[W/m$^{2}$], respectively.

Similarly, there are no significant differences in the fitting
parameters when considering confined and eruptive flares separately.

\subsection{Flare Reconnection Flux Ratio and Area Ratio in Confined and Eruptive Flares}

In Figures 5(a)-(b), we display the histograms of flare reconnection
flux ratio R$_{flux}$ ($\Phi$$_{ribbon}$/$\Phi$$_{AR}$) and ribbon
area ratio R$_{S}$ (S$_{ribbon}$/S$_{AR}$) for confined and eruptive
events. It can be seen that the distributions of both R$_{flux}$ and
R$_{S}$ show evident differences, with R$_{flux}$ and R$_{S}$ for
confined events smaller than those for eruptive flares. R$_{flux}$
ranges between 1\% and 41\% for eruptive flares and ranges between
1\% and 21\% for confined events. The proportion of eruptive flares
reaches $\sim$89\% (39 of 44) corresponding to the flux ratio
R$_{flux}$ higher than 15\%. The log averages of flux ratio
R$_{flux}$ are 6.3\% for confined and 9.5\% for eruptive events.
Similarly, the confined flares have the smaller area ratio R$_{S}$
(1\%$-$18\%) than eruptive events (1\%$-$30\%). The log-mean values
of area ratio R$_{S}$ are 4.0\% for confined and 6.1\% for eruptive
cases.

\section{Appearances of Typical Flare Examples}
\subsection{Two Confined Flares Within ARs of Large Magnetic Flux}

We investigate the dynamic evolution of confined flares originating
from ARs with a large magnetic flux ($\geq$ 1.0$\times$$10^{23}$
Mx), including 26 events from 5 different ARs (ARs 11339, 11520,
11967, 12192 and 12242). After examining the AIA observations of
these confined flares, we find that they have common
characteristics: slipping reconnection, strong shear, and a stable
filament. Here, two confined events from ARs 11520 and 12242 are
taken as examples to analyze the flare dynamics and magnetic
topological structures in detail.

On 10 July 2012, a confined M1.7 flare occurred in the sigmoidal
region of AR 11520 with $\Phi$$_{AR}$ of 1.24$\times$$10^{23}$ Mx.
The flare was initiated at 04:58 UT and the GOES SXR flux peaked at
05:14 UT. Before the flare started, a filament was located along the
PIL at the flaring region (left panel in Figure 6(a)). It did not
show any rise process during the flare and was stably present after
the flare (right panel in Figure 6(a)). The comparison of the 304
{\AA} image with the HMI LOS magnetogram showed that the flare
consisted of two positive-polarity ribbons PR1-PR2 and two
negative-polarity ribbons NR1-NR2 (middle panel in Figure 6(a)).
Ribbons PR1 and NR1 were located at two ends of the filament and PR2
and NR2 at both sides of the main body (axis) of the filament.
High-temperature flare loops at 94 {\AA} displayed notable dynamic
evolution (Figure 6(b)). To display the fine structures of the EUV
images, the 94 {\AA} filter channel data have been processed using
the multi-scale Gaussian normalization (MGN) method (Morgan \&
Druckm{\"u}ller 2014). At the start of the flare, two sets of loop
bundles L1 (connecting PR2-NR1) and L2 (connecting PR1-NR2)
overlying the stable filament became bright. Starting from about
05:10 UT, a group of brightened short loops L3 were formed
connecting PR2 and NR2, and meanwhile another longer loop bundles L4
linking PR1-NR1 can be discerned. During the flare, the north parts
of loop bundles L2 and L3 exhibited apparent bidirectional slipping
motions along ribbon NR2. Finally, strongly sheared post-flare loops
(PFLs) appeared above the non-eruptive filament. Based on the
dynamic evolution of flare loops and their relations with flare
ribbons, we suggest that slipping magnetic reconnection (Priest \&
D{\'e}moulin 1995; Aulanier et al. 2006) between loop bundles L1 and
L2 occurred and led to the formation of L3 and L4. We estimated the
inclination angles $\theta$ of PFLs with respect to the PIL,
corresponding to the angle between the tangents of the PFL and PIL
at their intersection (left panel in Figure 6(c)). The complementary
angle of $\theta$ has been referred to as the shear angle in
previous studies (Su et al. 2007; Aulanier et al. 2012). We derive
small $\theta$ values, ranging from 10$^{\circ}$ to 30$^{\circ}$,
indicative of a high non-potentiality in the form of a strong shear.
Then more high-temperature PFLs gradually cooled down and formed
PFLs overlying the stable filament at 171 {\AA} (Figure 6(c)).

Using the photospheric vector magnetic field observed by SDO/HMI at
04:24 UT, we make a nonlinear force-free (NLFFF) extrapolation by
applying the optimization method (Wheatland et al. 2000; Wiegelmann
2004) and obtain the 3D coronal magnetic fields. There are two sets
of sheared magnetic systems (MS1 and MS2 in Figure 7(a)) overlying a
twisted flux rope (FR) prior to the flare onset. By comparing the
AIA observations with the extrapolation results, we suggest that the
two magnetic systems MS1 and MS2 approximately correspond to two
sets of loop bundles L1 and L2 (Figure 6(b)) and the flux rope FR
bears a good resemblance to the observed non-eruptive filament
(Figure 6(a)). Based on the extrapolated 3D coronal magnetic field,
we calculated the squashing factor Q (Liu et al. 2016b) which
defines the locations of the quasi-separatrix layers (QSLs)
(D{\'e}moulin et al. 1996; Titov et al. 2002). As seen from the
distribution of Q (Figure 7(b)), the observed flare ribbons are
roughly matching the locations of high Q values, implying that
magnetic reconnection involved in the flare probably occurs in
regions of very strong magnetic connectivity gradients, i.e., QSLs.

Figure 8 shows another confined M1.3-class flare on 19 December 2014
in AR 12242, which has a large $\Phi$$_{AR}$ of $\sim$
1.11$\times$$10^{23}$ Mx. The flare was initiated at 09:31 UT and
peaked at 09:44 UT. It occurred at the northwest of AR 12242 and a
filament was present along the PIL at the flaring region (Figure
8(a)). During the flare process, the mainbody of the filament did
not show any rise phase except for the mild activation at its south
part. After the flare, the filament remained stabilized, similar to
the filament in the confined M1.7-class flare in AR 11520 (Figure
6(a)). Two quasi-parallel flare ribbons were distinguished from AIA
304 {\AA} images, including ribbon PR in the leading
positive-polarity sunspot and the negative-polarity ribbon NR. As
shown from the high-temperature 131 {\AA} observations, the flare
loops were composed of two sets of magnetic systems S1 and S2
overlying the non-eruptive filament, displaying a clear ``X-shape"
structure (Figure 8(b)). The south ends of systems S1 and S2 were
anchored in ribbon PR and their north ends in ribbon NR. During the
flare evolution, S1 and S2 exhibited apparent slipping motions along
ribbons PR and NR, and more flare loops successively appeared. In
the gradual phase of the flare, low-temperature PFLs were formed as
best observed in the AIA 171 {\AA} channel (Figure 8(c)). Similarly,
the early formed PFLs also displayed an ``X-shape" structure. We
estimated the inclination angles $\theta$ of PFLs with respect to
the PIL and found $\theta$ values in the range of
20$^{\circ}$$-$28$^{\circ}$. The small $\theta$ values imply that
the PFLs are strongly sheared and have a higher non-potentiality.

The apparent slipping motions of the fine structures within flare
ribbons are further displayed in Figure 9. Ribbon PR was composed of
numerous bright dot-like substructures, corresponding to the
footpoints of high-temperature flare loops. These substructures
exhibited apparent slipping motions in opposite directions (Figure
9(a)). We followed the trails of 3 different substructures within
ribbon PR. From 09:33:10 UT, the bright knot ``1" slipped toward the
east with a displacement of about 3.8 Mm in 110 s (with a velocity
of $\sim$30 km s$^{-1}$). Meanwhile, another bright knot ``2"
displayed a rapid slipping motion in the opposite direction at a
velocity of $\sim$130 km s$^{-1}$. At 09:34:58 UT, the bright knot
``3" at the middle part of PR underwent a fast slippage towards the
northeast. In order to analyze the slipping motions of the
substructures, we create a stack plot (Figure 9(c)) along slice
``C-D" in the AIA 131 {\AA} images (blue curve in Figure 9(a)). As
seen from the stack plot, the slipping motions along ribbon PR were
in both directions with speeds of 20$-$150 km s$^{-1}$. Figure 9(b)
shows the stack plot of the other ribbon NR along slice ``A-B"
(green dash-dotted curve in Figure 8(b)). Similarly, the slippage
along ribbon NR was bi-directional with apparent speeds of 20$-$30
km s$^{-1}$, smaller than those of ribbon PR.

\subsection{One Special Eruptive Event Within an AR of Large Magnetic Flux}

A large majority of flares (26 of 28 events) originating from ARs of
$\Phi$$_{AR}$ $\geq$ 1.0$\times$$10^{23}$ Mx are confined, however
there are two special eruptive flares among the 28 events. One event
is M4.0 flare on 24 October 2014 and the other is X1.8 event on 20
December 2014. The X1.8 flare caused a sympathetic eruption of a
nearby quiescent filament and generated a CME. Figure 10 displays
the appearance of the eruptive M4.0 flare on 24 October 2014. The
flare was initiated at 07:37 UT and peaked at 07:48 UT. It was
located at the southeast of AR 12192, far away from the main PIL of
the AR. The flare was triggered by a blow-out jet as seen in AIA 304
and 131 {\AA} images, and produced a CME at 08:12 UT observed by
LASCO/C2. It was suggested that the eruptive flare on the southern
border of the AR was close to neighboring open field regions
(Thalmann et al. 2015) and thus the jet successfully escaped from
the solar surface and formed a CME.

\subsection{One Eruptive Flare Within an AR of Small Magnetic Flux}

About 92\% events occurring in ARs with $\Phi$$_{AR}$ smaller than
3.0$\times$$10^{22}$ Mx are eruptive. Here, we present an eruptive
M3.9 event on 02 October 2011 as an example (Figure 11). The flare
originated from AR 11305 (N09$^{\circ}$, W12$^{\circ}$) with a
smaller $\Phi$$_{AR}$ of $\sim$ 1.67$\times$$10^{22}$. It started at
00:37 UT and reached its peak at 00:50 UT. A high-temperature flux
rope erupted towards the southwest in 131 {\AA} image. The angle of
separation between \emph{SDO} and \emph{STEREO B} on 02 Oct 2011 was
around 97$^{\circ}$. An erupting CME bubble can be observed at the
west limb in \emph{STEREO B}/EUVI 195 {\AA} image. Starting from
about 01:05 UT, an Earth-directed CME was observed by the COR1
coronagraph aboard \emph{STEREO B}.

\subsection{One Eruptive and One Confined Flares Within the same AR of Medium Magnetic Flux}

When $\Phi$$_{AR}$ is between 3.0$\times$$10^{22}$ and
1.0$\times$$10^{23}$ Mx, almost one half of flares are confined.
Here, we show two examples from the same AR 11429 on 06 March 2012.
At 07:52 UT, an eruptive M1.0-class flare occurred in AR 11429 with
$\Phi$$_{AR}$ of about 6.78$\times$$10^{22}$ Mx measured before the
flare onset (Figure 12). A reverse-S shaped filament was located
along the PIL. During the flare, the middle part of the filament
erupted and caused a CME. The AR was emerging persistently and the
magnetic flux increased to $\sim$ 7.96$\times$$10^{22}$ Mx at 12:00
UT. Then another confined M2.1-class flare occurred at 12:23 UT and
peaked at 12:41 UT (Figure 13). The filament in the AR did not erupt
except for the activation at the north part. A flux rope was
illuminated and started to rise at 12:26 UT as observed in 131 {\AA}
images. The rise lasted for about 10 min and ceased at 12:37 UT.
Then the flux rope seemed to stay at a certain height and faded away
gradually. The eruption of the flux rope failed and did not generate
any CME.

\section{Summary and Discussion}

In this work, we established the extensive database of
eruptive/confined large flares in the \emph{SDO} era (a total of 322
events including 170 eruptive and 152 confined cases). The
morphological properties of flaring ARs and the flare ribbons, and
their statistical relationships have been investigated. Our study
delivered the following main results.

1. We find that the total unsigned magnetic flux $\Phi$$_{AR}$ of
ARs plays an important role in governing the eruptive character of
flares, and the proportion of eruptive flares exhibits a strong
anti-correlation with $\Phi$$_{AR}$ (r$_{s}$=-0.97). About 93\%
flares originating from ARs with an unsigned magnetic flux larger
than 1.0$\times$$10^{23}$ Mx are confined, i.e., are not associated
with a CME. About 92\% events occurring in ARs with $\Phi$$_{AR}$
smaller than 3.0$\times$$10^{22}$ Mx are eruptive.

2. We also find a very high positive correlation (r$_{s}$=0.86)
empirical relation between critical decay index height $h_{crit}$
and $\Phi$$_{AR}$. This implies that ARs with a large magnetic flux
have a strong magnetic cage, which confines the eruption. This is
the first time that such a fundamental relation between the total AR
flux and the confinement properties for large flares is derived.

3. We find that the flare ribbon reconnection flux and flare ribbon
area are correlated with the peak X-ray flux. There are no
significant differences in the fitting parameters when considering
confined and eruptive flares separately. These findings are
consistent with previous studies (Veronig \& Polanec 2015;
Kazachenko et al. 2017; Tschernitz et al. 2018), while the obtained
correlation coefficients between flare reconnection flux
$\Phi$$_{ribbon}$ and flare peak X-ray flux F$_{SXR}$ are different,
probably due to different ranges of flare classes in different
statistical studies.

4. The ratio of the AR magnetic flux that is involved in the flare
reconnection process ranges between 1\% and 41\% for eruptive flares
and between 1\% and 21\% for confined events. Similarly, the
confined flares have the smaller area ratio R$_{S}$ (1\%$-$18\%)
than eruptive events (1\%$-$30\%).

5. By analyzing the dynamic evolution of 26 confined flares
occurring in ARs with $\Phi$$_{AR}$$\geq$1.0$\times$$10^{23}$ Mx, we
find that these flares have several characteristics in common:
stable filament, slipping magnetic reconnection and strongly sheared
PFLs, belonging to ``type I" flares as proposed in our previous work
(Li et al. 2019).

Our results show that the magnetic flux of ARs is a key parameter in
determining the eruptive character of large solar flares, and the
proportion of eruptive flares exhibits a strong anti-correlation
with $\Phi$$_{AR}$. The relation was first found in our work and has
never been revealed before. This means that the association rate of
flares and CMEs is decreasing with the increasing magnetic flux of
ARs. This finding is further supported by the high correlation
obtained between $\Phi$$_{AR}$ and the critical height for torus
instability. The ARs seem to be classified into three situations
according to their different magnetic properties: strong confinement
($\Phi$$_{AR}$$\geq$1.0$\times$$10^{23}$ Mx or $h_{crit}$$\geq$47
Mm), moderate confinement (3.0$\times$$10^{22}$
$<$$\Phi$$_{AR}$$<$1.0$\times$$10^{23}$ Mx or 27$<$$h_{crit}$$<$47
Mm) and weak confinement ($\Phi$$_{AR}$$\leq$3.0$\times$$10^{22}$ Mx
or $h_{crit}$$\leq$27 Mm). The values we use to discriminate between
classes, 3.0$\times$$10^{22}$ Mx and 1.0$\times$$10^{23}$ Mx, are
arbitrary. In the case of strong confinement, the flare energy and
associated magnetic reconfigurations are insufficient to break
through the overlying field even if it is an X-class flare (e.g. AR
12192), thus tend to generate confined flares (Guo et al. 2010;
Sarkar \& Srivastava 2018; Jing et al. 2018; Duan et al. 2019). On
the contrary, if the constraining effect of the background field is
so small, small disturbance in the lower corona can result in the
generation of a CME, which explains the high proportion of eruptive
flares originating from ARs with a small magnetic flux. When the
confinement of overlying magnetic cage is moderate, almost one half
of flares are confined (Area B in Figure 1(a), also see two examples
in Figures 12-13). This indicates that the overlying confinement and
AR non-potentiality (Falconer et al. 2002; Nindos \& Andrews 2004)
may jointly determine the class of the flare in the
moderate-confinement environment. Previous statistical studies have
shown that confined flares are often located much closer to the AR
centers where the strapping field is higher, whereas eruptive flares
occur at the periphery of an AR (Wang \& Zhang 2007; Baumgartner et
al. 2018).

Moreover, it is also found that the flux ratio R$_{flux}$ and area
ratio R$_{S}$ for confined flares are significantly smaller than
those for eruptive events. This result is similar to the statistical
result of Toriumi et al. (2017), who showed the parameter of the
ribbon area normalized by the sunspot area determines whether a
given flare is eruptive or not. They suggested that the relative
structural relation between the flaring region and the entire AR
controls the CME productivity.

Our findings reveal a new relation between the magnetic flux of ARs
and the occurrence of CMEs in association with large flares. They
also have important implications for stellar CMEs and the recently
detected ``superflares" on solar-type stars (Maehara et al. 2012;
Lynch et al. 2019). In order to produce a large flare, the magnetic
flux of the source AR has to be large (Aulanier et al. 2013; Shibata
et al. 2013; Tschernitz et al. 2018). The historical observational
data shows that the largest magnetic flux of flaring ARs is up to a
few times $10^{23}$ Mx (Zhang et al. 2010; Chen et al. 2011;
Schrijver et al. 2012). In the present study, between 1\% and 41\%
of the magnetic flux of the source AR for eruptive flares and
between 1\% and 21\% of $\Phi$$_{AR}$ for confined events are
involved in the flare reconnection process. If we assume that the
maximum percentage of 40\% of the magnetic flux contained in the AR
contributing to the flare reconnection process (Kazachenko et al.
2017; Tschernitz et al. 2018), a total reconnection flux of $\sim$
1.0$\times$$10^{23}$ Mx can be obtained for an AR with $\Phi$$_{AR}$
of 2.3$\times$$10^{23}$ Mx (the maximum magnetic flux in our
sample). According to the relation between the flare ribbon
reconnection flux and the peak X-ray flux (Equation 4 and Figure 4),
a flare of GOES class $\sim$ X100 could be powered.

Our results are interesting in two aspects: for the solar case, as
it means that for the strongest space weather effects (which are
predominantly due to the CME rather than the flare), we can not
simply extrapolate that the space weather effects will be
increasingly stronger for flares produced by large ARs present on
the Sun. Second, it has implications for the stellar case: we may
speculate that in case of the much larger ARs (stellar spots) that
are needed to produce the reported ``superflares" on solar-type
stars, the flares are probably mostly confined, as they will be
associated with a very strong overlying AR strapping field. This may
provide an explanation why the detection of stellar CMEs is rare
(Drake et al. 2013; Odert et al. 2017; Moschou et al. 2019;
Argiroffi et al. 2019), and implies that the solar-stellar
connection between flare rates and CME rates may be fundamentally
non-linear and actually ``breaking" when we come to the very large
events (Drake et al. 2013; Odert et al. 2017).

The confined flares from ARs with
$\Phi$$_{AR}$$\geq$1.0$\times$$10^{23}$ Mx are characterized by
slipping reconnection, strong shear, and a stable filament. They
belong to ``type I" confined flares proposed by Li et al. (2019),
who classified confined flares into two types based on their
different dynamic properties and magnetic configurations. Similar to
the appearance of confined flares in AR 12192 (Li et al. 2019), the
filaments in ARs 11339, 11520 (Figures 6-7), 11967 and 12242
(Figures 8-9) were all stably present and seemed not to be involved
in the flare evolution. The footpoints of high-temperature flare
loops exhibited apparent slipping motions in both directions along
flare ribbons (Figure 9), which implies the occurrence of slipping
magnetic reconnection overlying the non-eruptive filaments (Li \&
Zhang 2015; Dud{\'\i}k et al. 2016; L{\"o}rin{\v{c}}{\'\i}k et al.
2019; Shen et al. 2019; Chen et al. 2019). We suggest that slipping
flare loops along the two directions correspond to two different
magnetic systems, and the continuous slipping magnetic reconnection
between two magnetic systems causes the exchange of their magnetic
connectivity and apparent bi-directional slipping motions of
reconnecting field lines. Moreover, the PFLs observed in the gradual
phase of the flares were strongly sheared, indicating a high
non-potentiality. These observational characteristics of ``type I"
confined flares are inconsistent with the 2D standard CSHKP flare
model (Carmichael 1964; Sturrock 1966; Hirayama 1974; Kopp \&
Pneuman 1976), which suggests that the reconnection is associated
with the filament/flux rope eruption and occurs at a current sheet
below the erupting filament. Our observations have revealed several
different features. First, the filament/flux rope seemed to be
neither disturbed nor erupting during or after the flare. Second,
the reconnection site is more likely along the QSLs between two
magnetic systems overlying non-eruptive filaments. In summary, the
signatures of ``type I" confined flares in ARs with a large magnetic
flux pose a challenge to the 2D classical flare model and need to
establish 3D MHD models.

\acknowledgments {We thank the referee for helpful comments that
improved the paper. We thank Xudong Sun for useful discussions. This
work is supported by the National Natural Science Foundations of
China (11533008, 11773039, 11903050, 11673035, 11790304, 11673034,
11873059 and 11790300), the National Key R\&D Program of China
(2019YFA0405000), the B-type Strategic Priority Program of the
Chinese Academy of Sciences (XDB41000000), Key Programs of the
Chinese Academy of Sciences (QYZDJ-SSW-SLH050), Young Elite
Scientists Sponsorship Program by CAST (2018QNRC001), the Youth
Innovation Promotion Association of CAS (2014043 and 2017078) and
NAOC Nebula Talents Program. Lijuan Liu was supported by NSFC
(11803096) and the Open Project of CAS Key Laboratory of Geospace
Environment. Astrid M. Veronig acknowledges the support by the
Austrian Science Fund (FWF): P27292-N20. \emph{SDO} is a mission of
NASA's Living With a Star Program, \emph{STEREO} is the third
mission in NASA's Solar Terrestrial Probes Program, and \emph{SOHO}
is a mission of international cooperation between ESA and NASA.}

{}
\clearpage

\begin{figure}
\centering
\includegraphics
[bb=19 118 542 720,clip,angle=0,scale=0.8]{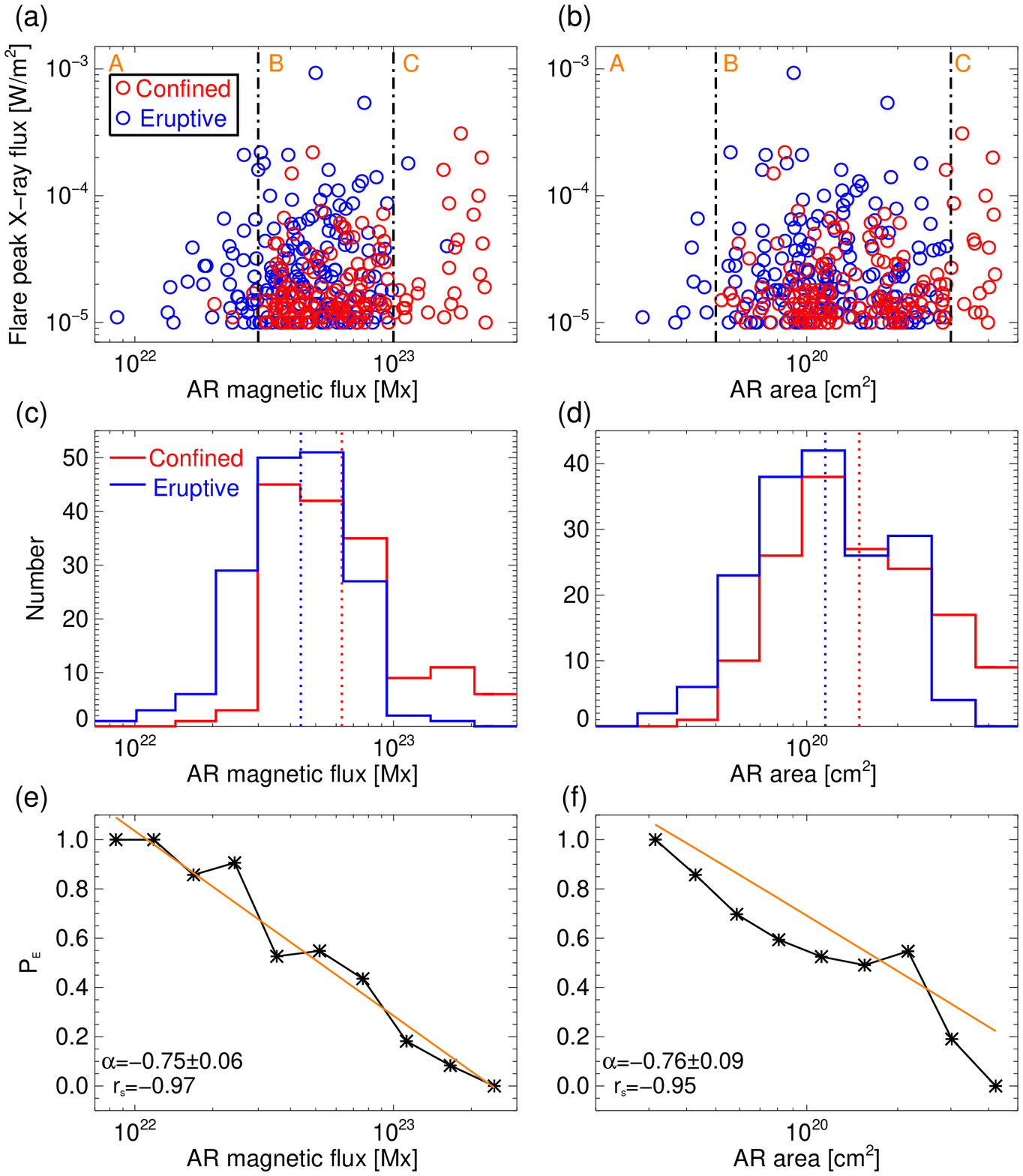}
\caption{Relations of the eruptive character of large solar flares
with unsigned AR magnetic flux and AR area. Top: scatter plots of
flare peak X-ray flux vs. unsigned AR magnetic flux and AR area.
Blue (red) circles are the eruptive (confined) flares. Two vertical
dash-dotted lines in panel (a) respectively refer to AR magnetic
flux of 3.0$\times$$10^{22}$ Mx and 1.0$\times$$10^{23}$ Mx. Two
vertical dash-dotted lines in panel (b) respectively correspond to
AR area of 5.0$\times$$10^{19}$ $cm^{2}$ and 3.0$\times$$10^{20}$
$cm^{2}$. Middle: histograms of AR magnetic flux and AR area for
confined (red) and eruptive (blue) events. Dotted vertical lines
indicate the means of the log values. Bottom: proportions of
eruptive flares P$_{E}$ vs. unsigned AR magnetic flux and AR area.
Orange lines show the results of linear fitting, and slopes $\alpha$
and Spearman rank order correlation coefficients r$_{s}$ are shown
at the bottom left. \label{fig1}}
\end{figure}
\clearpage

\begin{figure}
\centering
\includegraphics
[bb=15 304 554 525,clip,angle=0,scale=0.88]{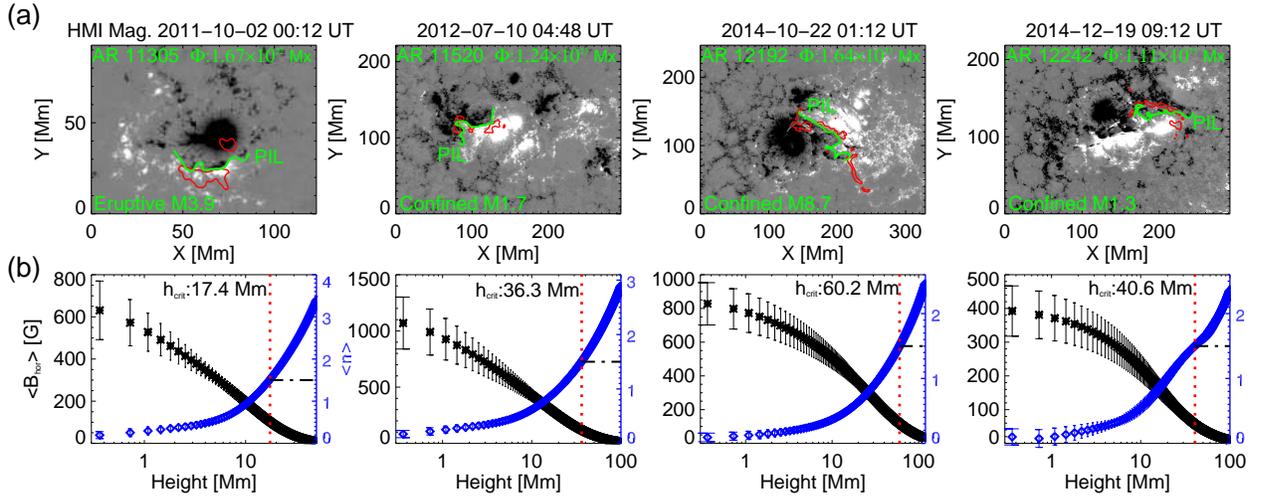} \caption{Decay
index of the coronal background fields for four examples. (a)
SDO/HMI photospheric magnetograms $B_{n}$ with contours of the AIA
1600 {\AA} flare ribbon brightenings (red contours) overplotted.
From left to right: eruptive M3.9-class flare in AR 11305, confined
flares of class M1.7 in AR 11520, M8.7 in AR 12192 and M1.3 in AR
12242. Green curves denote the flare-relevant PILs along which the
mean decay index is calculated. (b) $<$$B_{hor}$$>$ (black
asterisks) and $<$n$>$ (blue diamonds) as a function of height. The
error bars mark the corresponding standard deviation. Horizontal
black lines denote the position where $<$n$>$ equals 1.5 and red
vertical lines correspond to the critical height $h_{crit}$ for each
event. \label{fig2}}
\end{figure}
\clearpage

\begin{figure}
\centering
\includegraphics
[bb=20 317 531 516,clip,angle=0,scale=0.85]{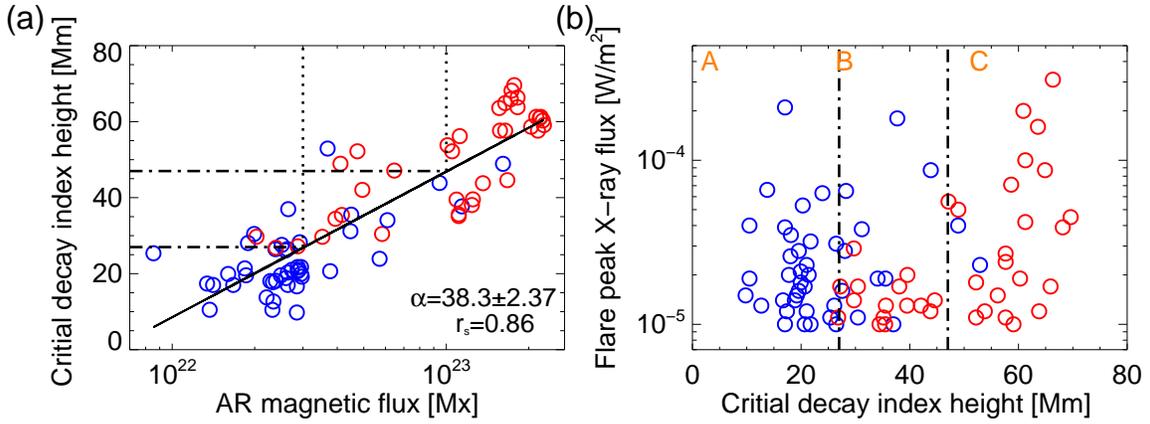} \caption{
Scatter plots of critical decay index height vs. unsigned AR
magnetic flux and flare peak X-ray flux vs. critical decay index
height. Blue (red) circles are the eruptive (confined) flares. The
black solid line in panel (a) shows the result of a linear fitting,
and slope $\alpha$ and Spearman rank order correlation coefficients
r$_{s}$ are shown at the bottom right. Two vertical dotted lines in
panel (a) denote the positions where $\Phi$$_{AR}$ respectively
equals 3.0$\times$$10^{22}$ Mx and 1.0$\times$$10^{23}$ Mx. Two
horizontal dash-dotted lines in panel (a) and two vertical
dash-dotted lines in panel (b) respectively refer to critical decay
index height of 27 Mm and 47 Mm. \label{fig3}}
\end{figure}
\clearpage

\begin{figure}
\centering
\includegraphics
[bb=23 312 539 518,clip,angle=0,scale=0.85]{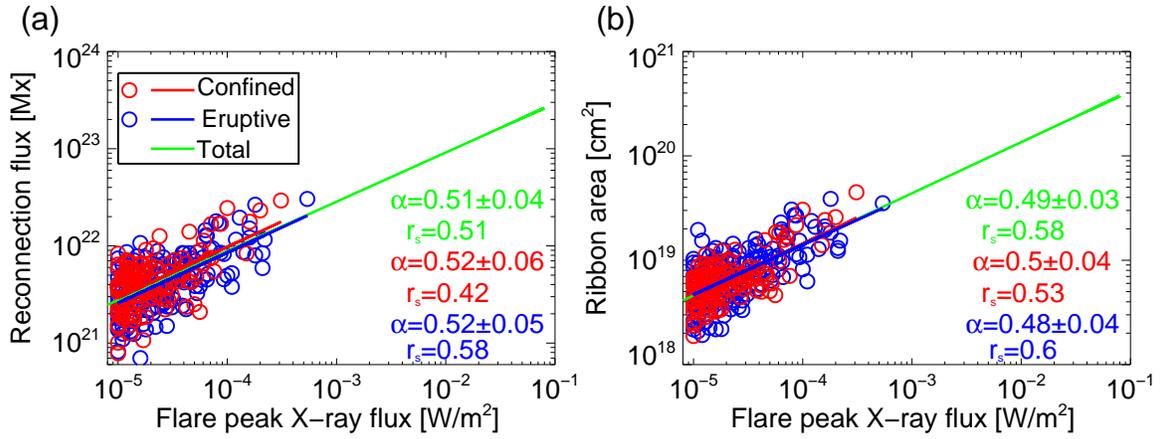} \caption{
Scatter plots of unsigned flare reconnection magnetic flux and
ribbon area vs. flare peak X-ray flux for confined (red) and
eruptive (blue) flares. Red, blue and green straight lines show the
results of linear fitting respectively for confined, eruptive and
total events. The slopes $\alpha$ and Spearman rank order
correlation coefficients r$_{s}$ are shown at the bottom right.
\label{fig4}}
\end{figure}
\clearpage

\begin{figure}
\centering
\includegraphics
[bb=39 315 516 513,clip,angle=0,scale=0.85]{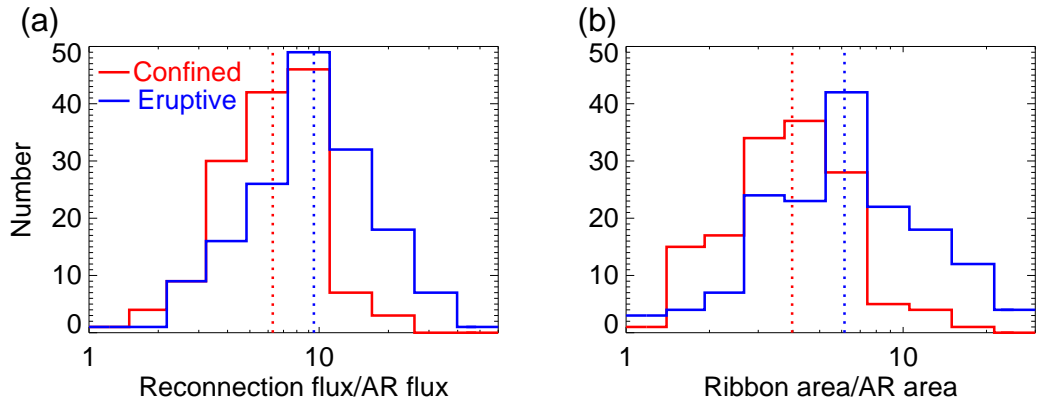}
\caption{Histograms of the ratios of flare reconnection flux/AR flux
and ribbon area/AR area for confined (red) and eruptive (blue)
events. Dotted vertical lines indicate the means of the log values.
\label{fig5}}
\end{figure}
\clearpage

\begin{figure}
\centering
\includegraphics
[bb=28 223 519 610,clip,angle=0,scale=0.85]{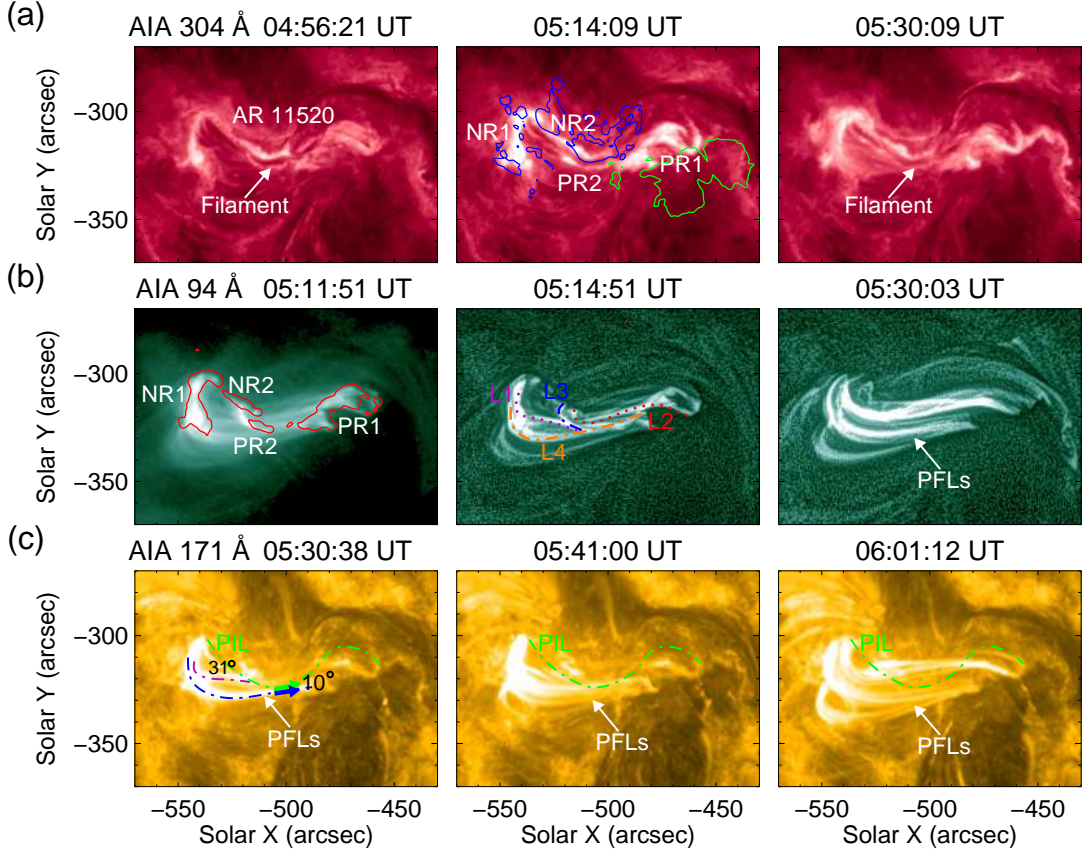} \caption{
Appearance of the confined M1.7-class flare in AR 11520 on 10 July
2012. (a) SDO/AIA 304 {\AA} images showing the stable filament
before and after the flare. Green and blue contours are the LOS
magnetic fields at $\pm$750 G levels. PR1-PR2 are two
positive-polarity flare ribbons and NR1-NR2 are two
negative-polarity ribbons. (b) AIA 94 {\AA} images displaying the
dynamic evolution of high-temperature flare loops. Red contours
denote the AIA 1600 {\AA} flare brightenings. L1-L4 outline four
sets of loop bundles and white arrow points to sheared post-flare
loops. (c) AIA 171 {\AA} images showing the low-temperature PFLs.
Two sets of PFLs (purple and blue dashed-dotted curves) are
delineated to estimate their inclination angles with respect to the
PIL (green dash-dotted line). The animation of this figure includes
AIA 304, 171 and 94 {\AA} images from 04:58 UT to 05:41 UT. The
video duration is 3 s.\label{fig6}}
\end{figure}
\clearpage

\begin{figure}
\centering
\includegraphics
[bb=22 314 574 520,clip,angle=0,scale=0.8]{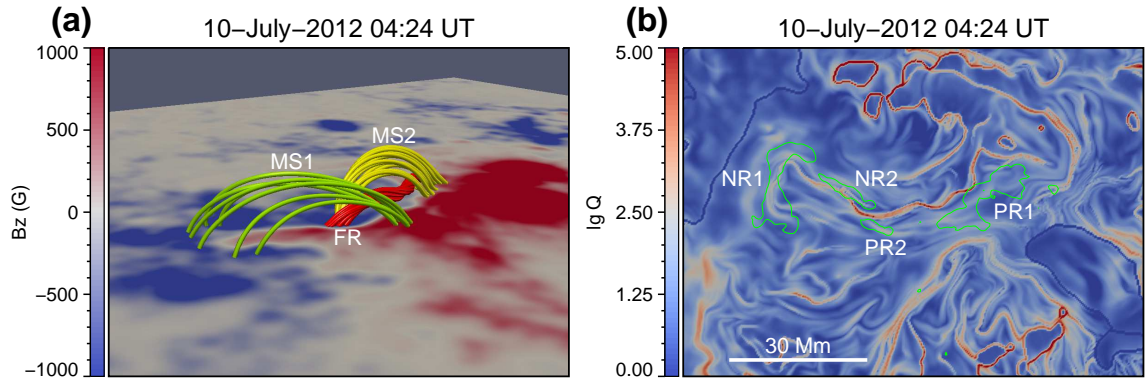}
\caption{Magnetic field configuration of the flare region. (a) Side
view of extrapolated field lines showing two magnetic systems
MS1-MS2 and the underlying flux rope FR. (b) Map of the squashing
factor Q on the HMI bottom boundary calculated from the nonlinear
force-free field. Green contours outline the flare ribbon
brightenings in the AIA 1600 {\AA} channel. \label{fig7}}
\end{figure}
\clearpage

\begin{figure}
\centering
\includegraphics
[bb=30 227 538 603,clip,angle=0,scale=0.9]{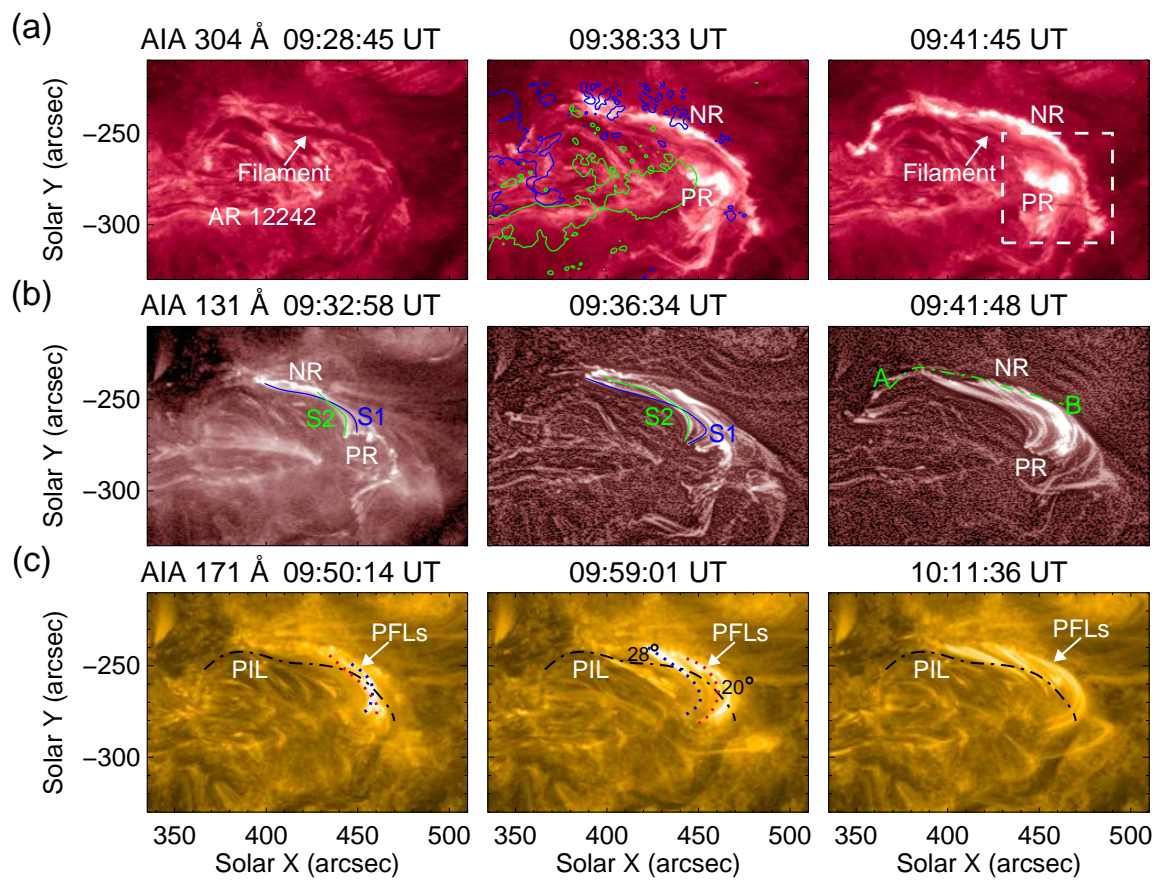}
\caption{Appearance of the confined M1.3-class flare in AR 12242 on
19 December 2014. (a) SDO/AIA 304 {\AA} images showing the
non-eruptive filament throughout the flare. Green and blue contours
are the HMI LOS magnetic fields at $\pm$350 G levels. NR and PR are
two quasi-parallel flare ribbons. The white square denotes the field
of view (FOV) of Figure 9. (b) AIA 131 {\AA} images displaying two
magnetic systems S1 and S2. Green dash-dotted curve ``A-B" shows the
cut position used to obtain the stack plot shown in Figure 9. (c)
AIA 171 {\AA} images showing the low-temperature PFLs. Red and blue
dotted curves outline the PFLs at different times and the black
dash-dotted line delineates the PIL associated with the flare. The
animation of this figure includes AIA 304, 171 and 131 {\AA} images
from 09:25 UT to 10:11 UT. The video duration is 3 s. \label{fig8}}
\end{figure}
\clearpage

\begin{figure}
\centering
\includegraphics
[bb=30 120 654 700,clip,angle=0,scale=0.8]{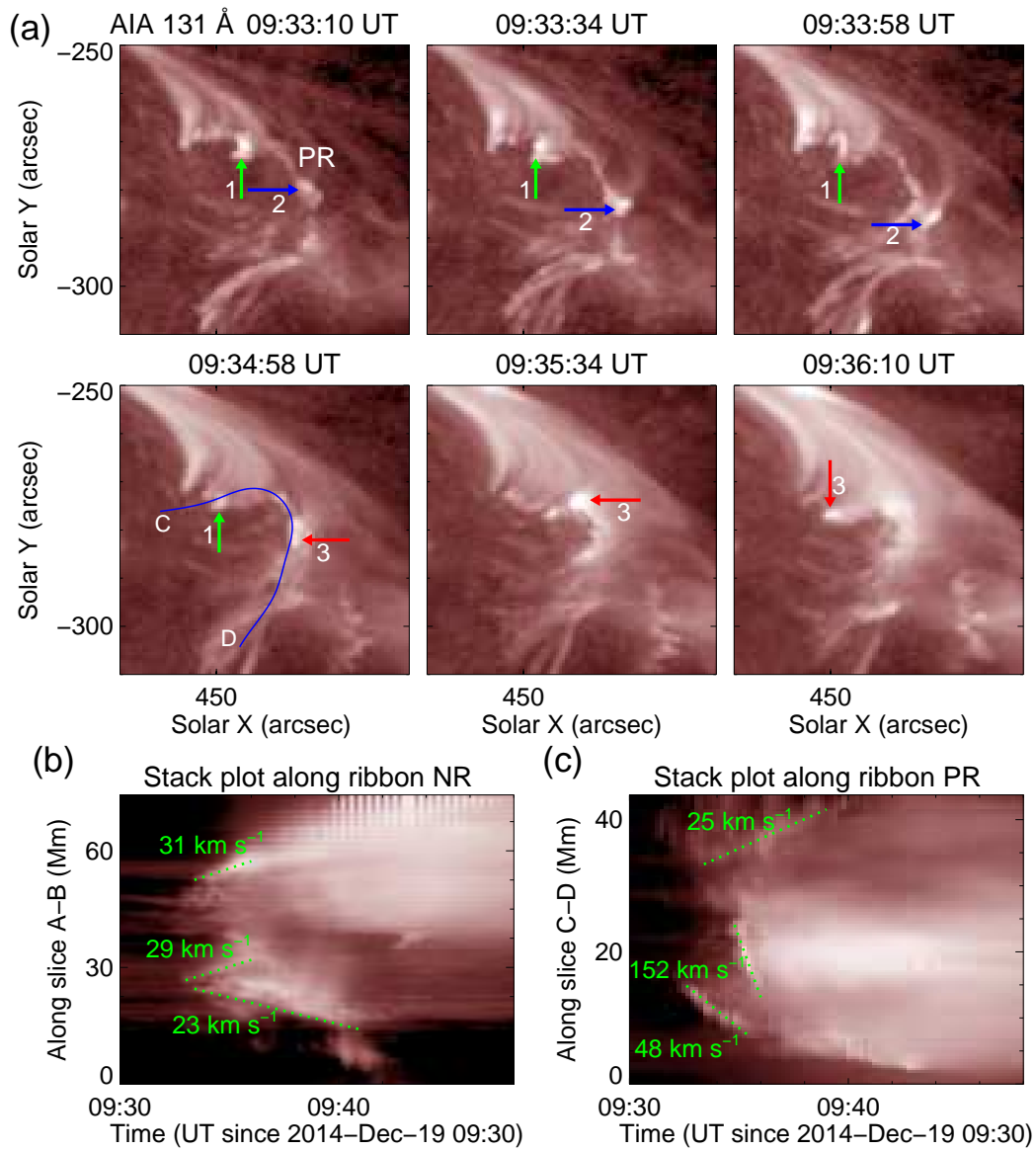}
\caption{Apparent slipping motions of fine structures within flare
ribbons PR and NR. (a) Time series of SDO/AIA 131 {\AA} images
showing the slippage of traced bright knots (``1"$-$``3") within
ribbon PR. Bright knots ``1" and ``3" slipped toward the east end of
PR and knot ``2" slipped in the opposite direction. The blue curve
``C-D" shows the cut position used to obtain the stack plot shown in
panel (c). (b)-(c) 131 {\AA} stack plots along slices ``A-B" and
``C-D" showing the bidirectional slippage along ribbons NR and PR.
\label{fig9}}
\end{figure}
\clearpage

\begin{figure}
\centering
\includegraphics
[bb=17 175 531 664,clip,angle=0,scale=0.8]{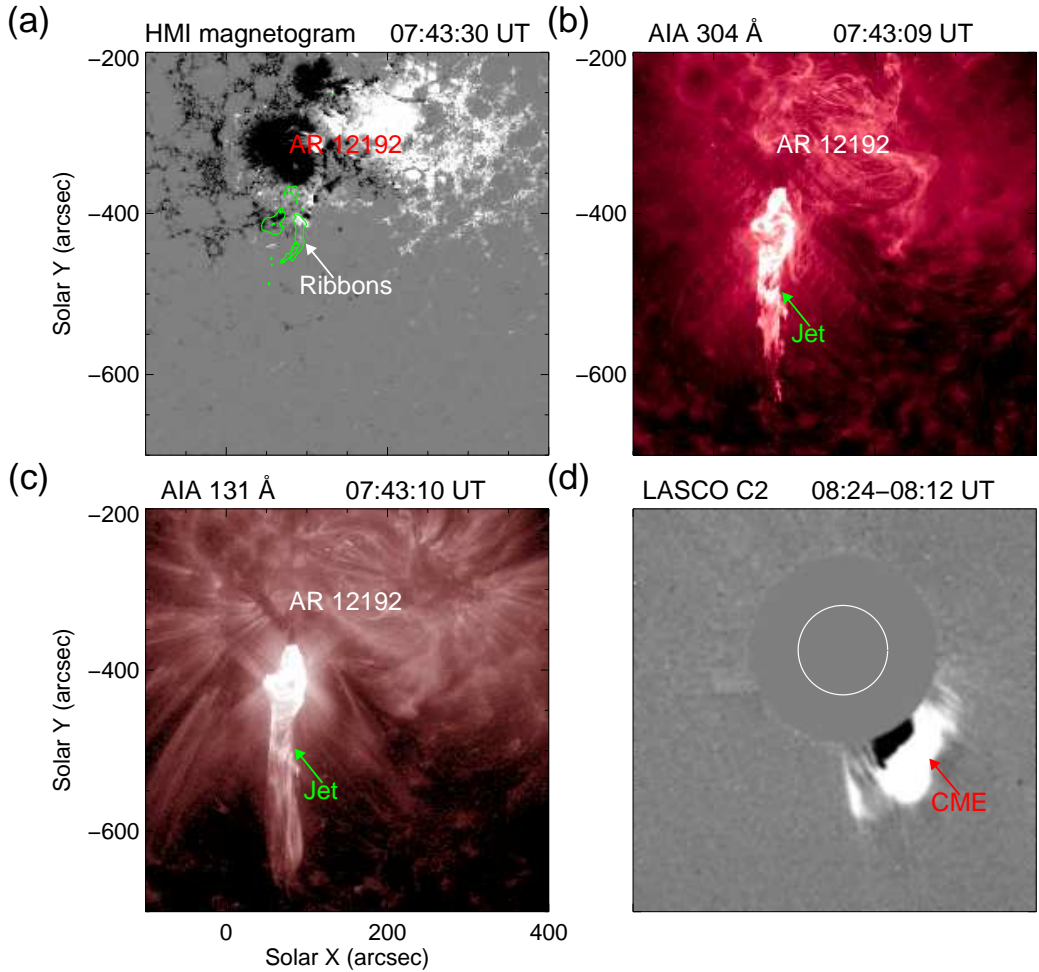}
\caption{Appearance of the eruptive M4.0-class flare in AR 12192 on
24 October 2014. (a) SDO/HMI LOS magnetogram with contours of the
AIA 1600 {\AA} flare brightenings (green contours) overplotted.
(b)-(c) AIA 304 and 131 {\AA} images displaying the blow-out jet at
the southeast edge of AR 12192. (d) LASCO/C2 running-difference
image showing the associated CME. \label{fig10}}
\end{figure}
\clearpage

\begin{figure}
\centering
\includegraphics
[bb=30 163 533 676,clip,angle=0,scale=0.8]{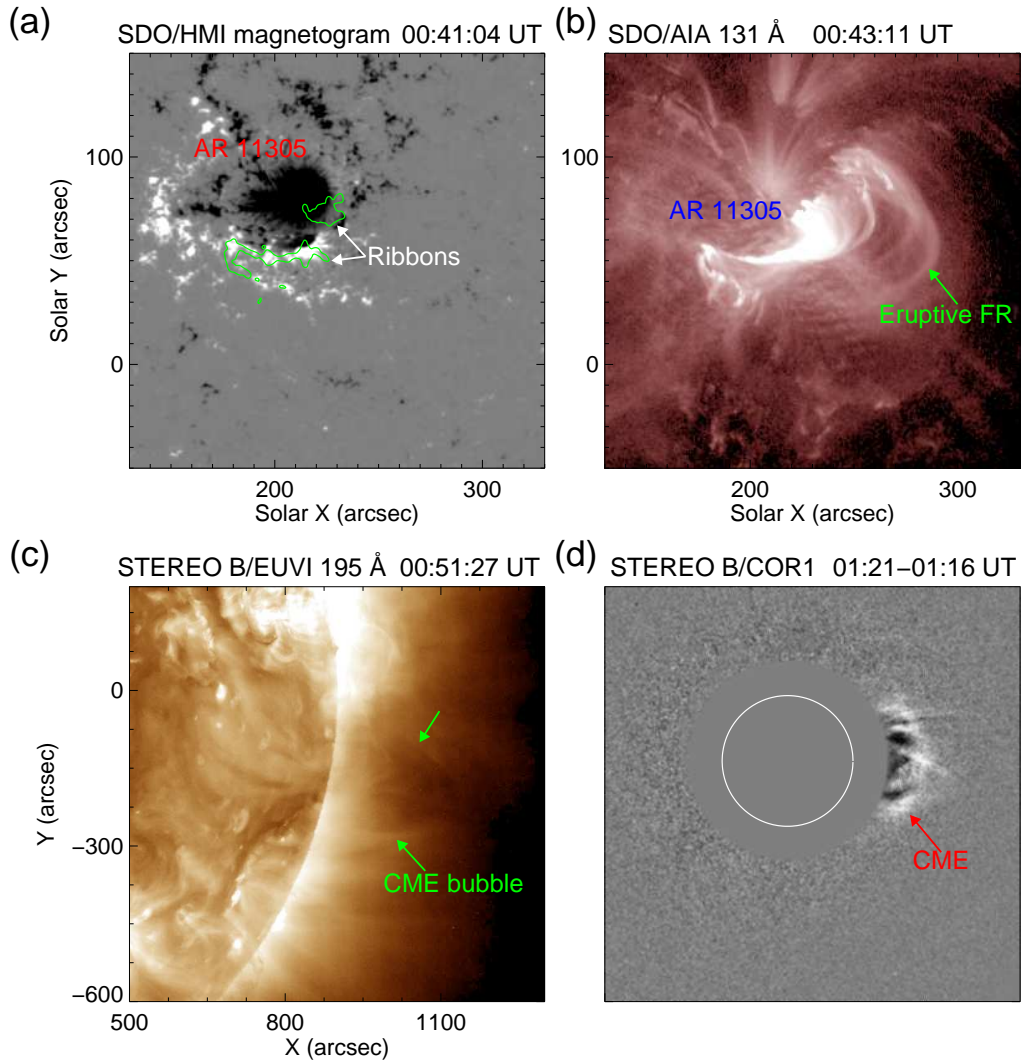}
\caption{Overview of the eruptive M3.9 flare on 02 October 2011 in
AR 11305. (a) SDO/HMI LOS magnetogram with contours of the AIA 1600
{\AA} flare brightenings (green contours) overplotted. (b) AIA 131
{\AA} image displaying the eruptive flux rope. (c) STEREO B/EUVI 195
{\AA} image showing the generated CME bubble. (d) STEREO B/COR1
image displaying the associated CME. \label{fig11}}
\end{figure}
\clearpage

\begin{figure}
\centering
\includegraphics
[bb=24 175 528 663,clip,angle=0,scale=0.8]{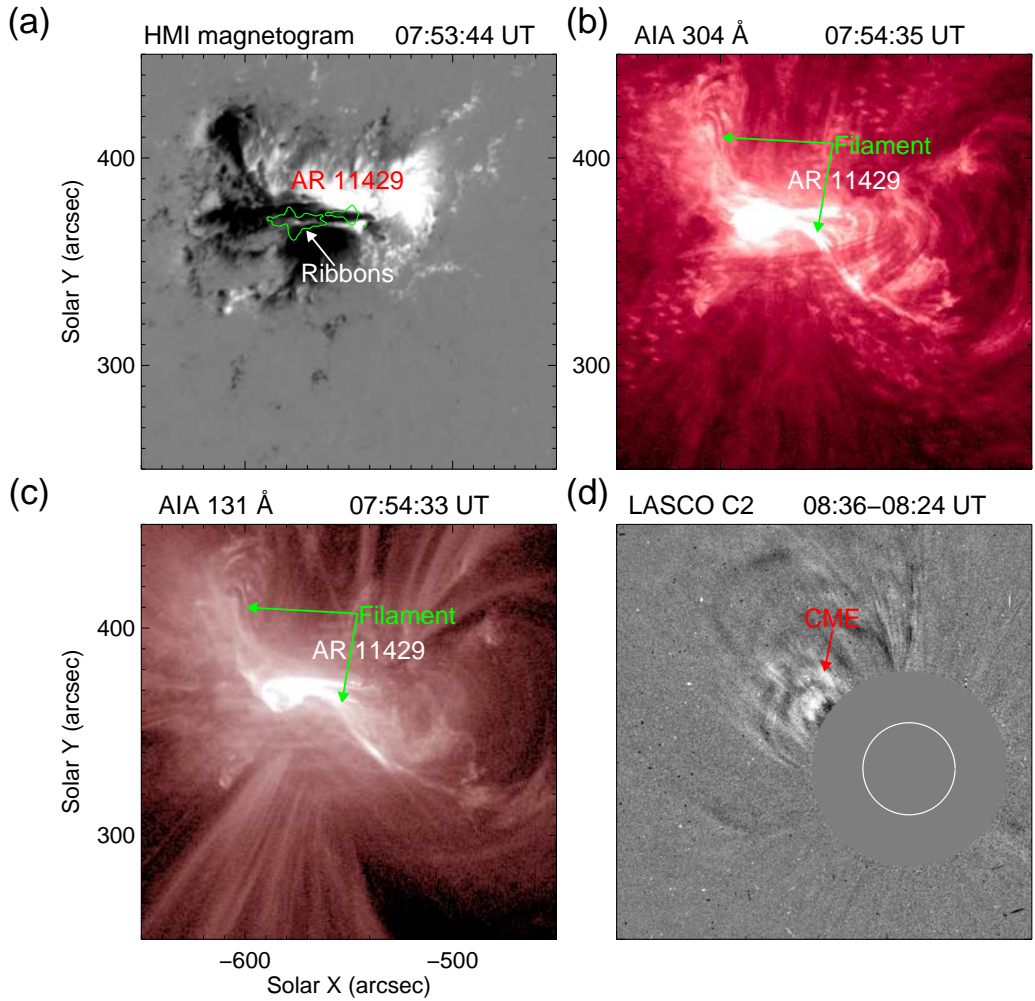}
\caption{Appearance of the eruptive M1.0-class flare in AR 11429 on
06 March 2012. (a) SDO/HMI LOS magnetogram with contours of the AIA
1600 {\AA} flare brightenings (green contours) overplotted. (b)-(c)
AIA 304 and 131 {\AA} images displaying the partial-eruptive
filament. (d) LASCO/C2 running-difference image showing the
associated CME. \label{fig12}}
\end{figure}
\clearpage

\begin{figure}
\centering
\includegraphics
[bb=24 175 528 663,clip,angle=0,scale=0.8]{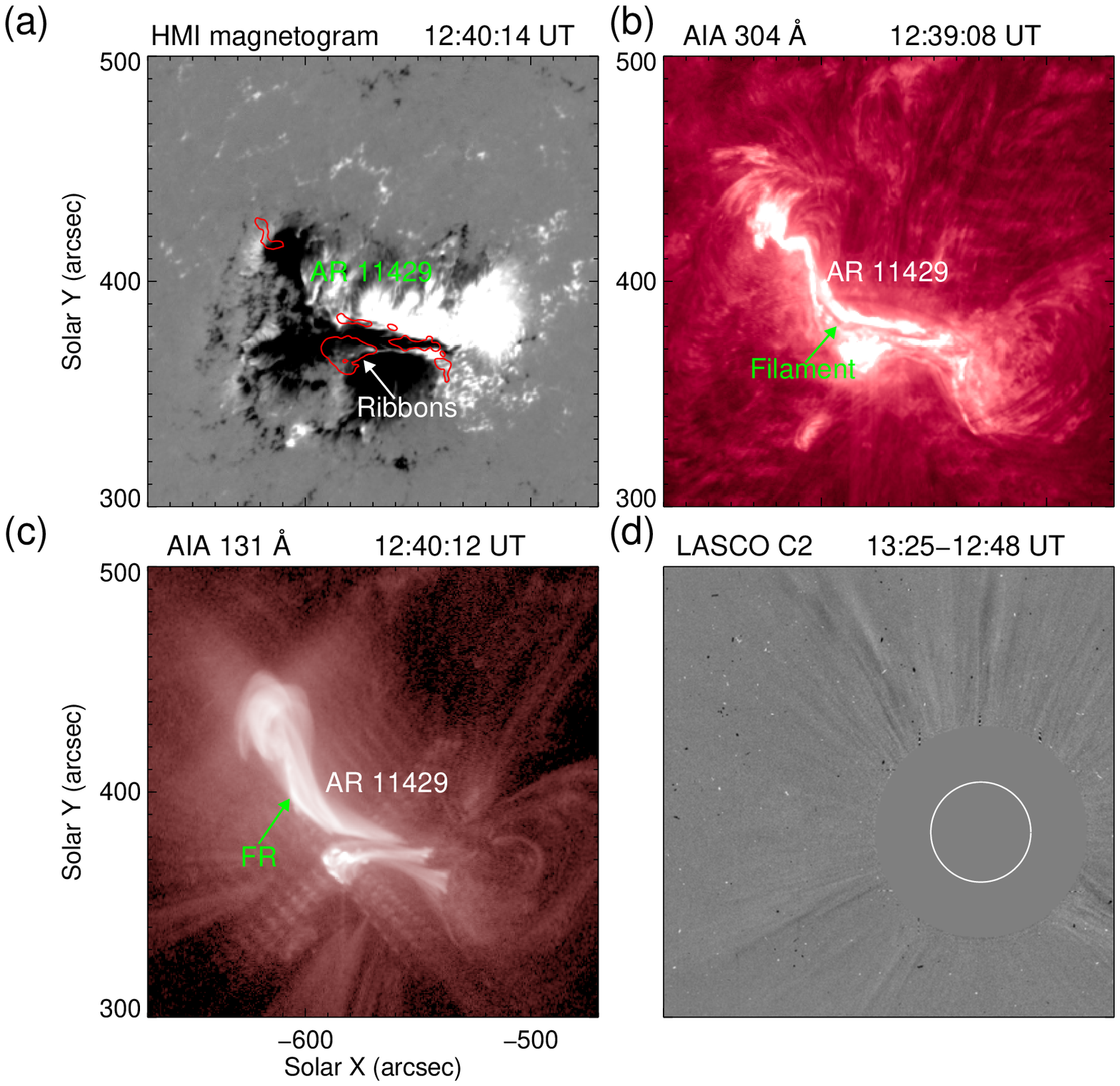}
\caption{Appearance of the confined M2.1-class flare in AR 11429 on
06 March 2012. (a) SDO/HMI LOS magnetogram with contours of the AIA
1600 {\AA} flare brightenings (red contours) overplotted. (b) AIA
304 {\AA} image displaying the non-eruptive filament along the PIL.
(c) AIA 131 {\AA} image showing the failed eruption of the flux
rope. (d) LASCO/C2 running-difference image. \label{fig13}}
\end{figure}
\clearpage


\begin{thebibliography}{}
\bibitem[()]{}
Alissandrakis, C.~E.\ 1981, \aap, 100, 197
\bibitem[()]{}
Amari, T., Canou, A., Aly, J.-J., Delyon, F., \& Alauzet, F.\ 2018,
\nat, 554, 211
\bibitem[()]{}
Andrews, M.~D.\ 2003, \solphys, 218, 261
\bibitem[()]{}
Argiroffi, C., Reale, F., Drake, J.~J., et al.\ 2019, Nature
Astronomy, 3, 742
\bibitem[()]{}
Aulanier, G., D{\'e}moulin, P., Schrijver, C.~J., et al.\ 2013,
\aap, 549, A66
\bibitem[()]{}
Aulanier, G., Janvier, M., \& Schmieder, B.\ 2012, \aap, 543, A110
\bibitem[()]{}
Aulanier, G., Pariat, E., D{\'e}moulin, P., \& DeVore, C.~R.\ 2006,
\solphys, 238, 347
\bibitem[()]{}
Baumgartner, C., Thalmann, J.~K., \& Veronig, A.~M.\ 2018, \apj,
853, 105
\bibitem[()]{}
Bobra, M.~G., Sun, X., Hoeksema, J.~T., et al.\ 2014, \solphys, 289,
3549
\bibitem[()]{}
Bokenkamp, N. 2007, PhD thesis, Stanford Univ.
\bibitem[()]{}
Carmichael, H.\ 1964, NASA Special Publication, 50, 451
\bibitem[()]{}
Chen, A.~Q., Wang, J.~X., Li, J.~W., et al.\ 2011, \aap, 534, A47
\bibitem[()]{}
Chen, H., Yang, J., Ji, K., et al.\ 2019, \apj, 887, 118
\bibitem[()]{}
Chen, H., Zhang, J., Ma, S., et al.\ 2015, \apjl, 808, L24
\bibitem[()]{}
Cheng, X., Zhang, J., Ding, M.~D., Guo, Y., \& Su, J.~T.\ 2011,
\apj, 732, 87
\bibitem[()]{}
D{\'e}moulin, P., Priest, E.~R., \& Lonie, D.~P.\ 1996, \jgr, 101,
7631
\bibitem[()]{}
Drake, J.~J., Cohen, O., Yashiro, S., et al.\ 2013, \apj, 764, 170
\bibitem[()]{}
Duan, A., Jiang, C., He, W., et al.\ 2019, \apj, 884, 73
\bibitem[()]{}
Dud{\'{\i}}k, J., Polito, V., Janvier, M., et al.\ 2016, \apj, 823,
41
\bibitem[()]{}
Falconer, D.~A., Moore, R.~L., \& Gary, G.~A.\ 2002, \apj, 569, 1016
\bibitem[()]{}
Fan, Y., \& Gibson, S.~E.\ 2007, \apj, 668, 1232
\bibitem[()]{}
Forbes, T.~G.\ 2000, \jgr, 105, 23153
\bibitem[()]{}
Gopalswamy, N., Yashiro, S., Michalek, G., et al.\ 2009, Earth Moon
and Planets, 104, 295
\bibitem[()]{}
Gosling, J.~T., McComas, D.~J., Phillips, J.~L., et al.\ 1991, \jgr,
96, 7831
\bibitem[()]{}
Green, L.~M., Matthews, S.~A., van Driel-Gesztelyi, L., Harra,
L.~K., \& Culhane, J.~L.\ 2002, \solphys, 205, 325
\bibitem[()]{}
Green, L.~M., T{\"o}r{\"o}k, T., Vr{\v{s}}nak, B., et al.\ 2018,
\ssr, 214, 46
\bibitem[()]{}
Guo, Y., Ding, M.~D., Schmieder, B., et al.\ 2010, \apjl, 725, L38
\bibitem[()]{}
Hagyard, M.~J., \& Rabin, D.~M.\ 1986, Advances in Space Research,
6, 7
\bibitem[()]{}
Hirayama, T.\ 1974, \solphys, 34, 323
\bibitem[()]{}
Howard, R.~A., Moses, J.~D., Vourlidas, A., et al.\ 2008, \ssr, 136,
67
\bibitem[()]{}
Jing, J., Liu, C., Lee, J., et al.\ 2018, \apj, 864, 138
\bibitem[()]{}
Kaiser, M.~L., Kucera, T.~A., Davila, J.~M., et al.\ 2008, \ssr,
136, 5
\bibitem[()]{}
Kazachenko, M.~D., Lynch, B.~J., Welsch, B.~T., et al.\ 2017, \apj,
845, 49
\bibitem[()]{}
Khodachenko, M.~L., Ribas, I., Lammer, H., et al.\ 2007,
Astrobiology, 7, 167
\bibitem[()]{}
Kliem, B., \& T{\"o}r{\"o}k, T.\ 2006, \prl, 96, 255002
\bibitem[()]{}
Kopp, R.~A., \& Pneuman, G.~W.\ 1976, \solphys, 50, 85
\bibitem[()]{}
Lammer, H., Lichtenegger, H.~I.~M., Kulikov, Y.~N., et al.\ 2007,
Astrobiology, 7, 185
\bibitem[()]{}
Lemen, J. R., Title, A. M., Akin, D. J., et al. 2012, Sol. Phys.,
275, 17
\bibitem[()]{}
Li, T., Liu, L., Hou, Y., et al.\ 2019, \apj, 881, 151
\bibitem[()]{}
Li, T., \& Zhang, J.\ 2015, \apjl, 804, L8
\bibitem[()]{}
Liu, L., Wang, Y., Wang, J., et al.\ 2016a, \apj, 826, 119
\bibitem[()]{}
Liu, L., Wang, Y., Zhou, Z., et al.\ 2018, \apj, 858, 121
\bibitem[()]{}
Liu, R., Kliem, B., Titov, V.~S., et al.\ 2016b, \apj, 818, 148
\bibitem[()]{}
L{\"o}rin{\v{c}}{\'\i}k, J., Dud{\'\i}k, J., \& Aulanier, G.\ 2019,
\apj, 885, 83
\bibitem[()]{}
Lynch, B.~J., Airapetian, V.~S., DeVore, C.~R., et al.\ 2019, \apj,
880, 97
\bibitem[()]{}
Maehara, H., Shibayama, T., Notsu, S., et al.\ 2012, \nat, 485, 478
\bibitem[()]{}
Moore, R.~L., Sterling, A.~C., Hudson, H.~S., et al.\ 2001, \apj,
552, 833
\bibitem[()]{}
Morgan, H., \& Druckm{\"u}ller, M.\ 2014, \solphys, 289, 2945
\bibitem[()]{}
Moschou, S.-P., Drake, J.~J., Cohen, O., et al.\ 2019, \apj, 877,
105
\bibitem[()]{}
Nindos, A., \& Andrews, M.~D.\ 2004, \apjl, 616, L175
\bibitem[()]{}
Odert, P., Leitzinger, M., Hanslmeier, A., et al.\ 2017, \mnras,
472, 876
\bibitem[()]{}
Pesnell, W. D., Thompson, B. J., \& Chamberlin, P. C. 2012, Sol.
Phys., 275, 3
\bibitem[()]{}
Priest, E.~R., \& D{\'e}moulin, P.\ 1995, \jgr, 100, 23443
\bibitem[()]{}
Priest, E., \& Forbes, T.\ 2000, Magnetic Reconnection
\bibitem[()]{}
Sarkar, R., \& Srivastava, N.\ 2018, \solphys, 293, 16
\bibitem[()]{}
Scherrer, P.~H., Schou, J., Bush, R.~I., et al.\ 2012, \solphys,
275, 207
\bibitem[()]{}
Schrijver, C.~J., Beer, J., Baltensperger, U., et al.\ 2012, Journal
of Geophysical Research (Space Physics), 117, A08103
\bibitem[()]{}
Shen, Y., Qu, Z., Zhou, C., et al.\ 2019, \apjl, 885, L11
\bibitem[()]{}
Shibata, K., Isobe, H., Hillier, A., et al.\ 2013, \pasj, 65, 49
\bibitem[()]{}
Shibata, K., \& Magara, T.\ 2011, Living Reviews in Solar Physics,
8, 6
\bibitem[()]{}
Sturrock, P.~A.\ 1966, \nat, 211, 695
\bibitem[()]{}
Su, Y., Golub, L., \& Van Ballegooijen, A.~A.\ 2007, \apj, 655, 606
\bibitem[()]{}
Su, Y., Veronig, A.~M., Holman, G.~D., et al.\ 2013, Nature Physics,
9, 489
\bibitem[()]{}
Sun, X.\ 2013, arXiv e-prints, arXiv:1309.2392
\bibitem[()]{}
Sun, X., Bobra, M.~G., Hoeksema, J.~T., et al.\ 2015, \apjl, 804,
L28
\bibitem[()]{}
Svestka, Z.\ 1986, The Lower Atmosphere of Solar Flares, 332
\bibitem[()]{}
Thalmann, J.~K., Su, Y., Temmer, M., et al.\ 2015, \apjl, 801, L23
\bibitem[()]{}
Titov, V.~S., Hornig, G., \& D{\'e}moulin, P.\ 2002, Journal of
Geophysical Research (Space Physics), 107, 1164
\bibitem[()]{}
Toriumi, S., Schrijver, C.~J., Harra, L.~K., et al.\ 2017, \apj,
834, 56
\bibitem[()]{}
Tschernitz, J., Veronig, A.~M., Thalmann, J.~K., et al.\ 2018, \apj,
853, 41
\bibitem[()]{}
Vasantharaju, N., Vemareddy, P., Ravindra, B., et al.\ 2018, \apj,
860, 58
\bibitem[()]{}
Veronig, A.~M., \& Polanec, W.\ 2015, \solphys, 290, 2923
\bibitem[()]{}
Wang, Y., \& Zhang, J.\ 2007, \apj, 665, 1428
\bibitem[()]{}
Wheatland, M.~S., Sturrock, P.~A., \& Roumeliotis, G.\ 2000, \apj,
540, 1150
\bibitem[()]{}
Wiegelmann, T.\ 2004, \solphys, 219, 87
\bibitem[()]{}
Yang, S., Zhang, J., \& Xiang, Y.\ 2014, \apjl, 793, L28
\bibitem[()]{}
Yashiro, S., Akiyama, S., Gopalswamy, N., et al.\ 2006, \apjl, 650,
L143
\bibitem[()]{}
Zhang, J., Wang, Y., \& Liu, Y.\ 2010, \apj, 723, 1006
\bibitem[()]{}
Zuccarello, F.~P., Aulanier, G., \& Gilchrist, S.~A.\ 2015, \apj,
814, 126

\end{thebibliography}
\end{document}